\documentclass[twocolumn,floatfix,
showpacs,showkeys,footinbib,prl]{revtex4-2} 
\usepackage{epsfig}
\usepackage{latexsym}
\usepackage{xspace}
\usepackage[colorlinks=true,linktocpage=true,linkcolor=blue,citecolor=blue,allcolors=blue]{hyperref}
\usepackage{url}
\usepackage[utf8]{inputenc}
\usepackage{indentfirst}
\usepackage{enumerate}
\usepackage{color}

\usepackage[caption=false,position=top]{subfig}

\usepackage{amsmath}
\usepackage{amssymb}
\usepackage{url}

\newcommand{\supmat}{supplemental material}

\newcommand{\bs}{\boldsymbol}

\newcommand{\bvar}{\bs}
\newcommand{\mean}[1]{\left\langle #1 \right\rangle}

\newcommand{\eq}[1]{\begin{align} #1 \end{align}}
\newcommand{\rr}{{\bf r}}

\newcommand{\etam}{\eta_{\rm max}}
\newcommand{\sNN}{\sqrt{s_{\rm NN}}}

\newcommand{\cc}[1]{\chi_{#1}}

\begin{document}


\title{
Density correlations under global and local charge conservation
}

\author{Volodymyr Vovchenko}
\affiliation{Physics Department, University of Houston, 3507 Cullen Blvd, Houston, TX 77204, USA}

\begin{abstract}
This work presents a formalism to compute spatial $n$-point correlations of a conserved charge density in a large thermal system in the canonical ensemble, with explicit results presented up to 4th order.
The resulting correlators contain local and balancing terms expressed through the grand-canonical susceptibilities for any equation of state.
The new formalism is used to introduce a Gaussian local baryon number conservation in (spatial) rapidity in heavy-ion collisions at LHC conditions through a modulation of the balancing term.
Connection to the conservation volume $V_c$ approach is established, indicating that the latter is appropriate for observables within a limited range around midrapidity.
Quantitative analysis of the experimental data of the ALICE Collaboration on net proton cumulants in central Pb-Pb collisions at $\sNN = 5.02$~TeV shows significant evidence for local baryon conservation.
\end{abstract}


\maketitle


\paragraph*{\bf Introduction.}

Fluctuations and correlations of particle numbers produced in heavy-ion collisions are key observables in studies of the QCD phase structure, including the search for QCD critical point within energy scan programs at RHIC~\cite{Stephanov:1999zu,Luo:2017faz,Bzdak:2019pkr,STAR:2020tga,STAR:2021iop}, SPS~\cite{Gazdzicki:2015ska}, and FAIR/GSI~\cite{CBM:2016kpk,Almaalol:2022xwv,HADES:2020wpc}, and chiral criticality at the LHC~\cite{Friman:2011pf,ALICE:2022wwr}.
In the grand-canonical limit, fluctuations of conserved charges are directly proportional to the susceptibilities: $\kappa_n[B] = V \chi_n$, where $\chi_n = T^{n-1} \partial^n p / \partial \mu^n$~\footnote{The temperature variable $T$ has been dropped for brevity and the susceptibilities have the dimensions of the charge density. The dimensionless susceptibilities studied in lattice QCD are obtained by multiplying by $T^3$: $\chi_n \to T^3 \chi_n$. }.
The susceptibilities also determine the local density fluctuations, namely~\cite{landau2013statistical}
\eq{
\mean{\delta \rho (\mathbf{r}_1) \delta \rho (\mathbf{r}_2)}_{\rm local} \propto \chi_2 \, \delta(\rr_1 - \rr_2),
}
where $\delta(\rr_1 - \rr_2)$ should be understood as a localized function of the order of correlation length, $|\rr_1 - \rr_2| \lesssim \xi$.

However, the total charge is conserved in heavy-ion collisions, $\mean{\delta B_{\rm tot}^2} = \int d \rr_1 d \rr_2 \mean{\delta \rho (\mathbf{r}_1) \delta \rho (\mathbf{r}_2)}  = 0$, therefore, a long-range contribution to $\mean{\delta \rho (\mathbf{r}_1) \delta \rho (\mathbf{r}_2)}$ that balances the local part must exist~\cite{Bass:2000az,Pruneau:2002yf,Begun:2004gs,Ling:2013ksb}.  This is also the case for high-order, $n$-point charge density correlations.
Their possible non-equilibrium evolution on a hydrodynamics background has been discussed recently in several works~\cite{Kapusta:2011gt,Kapusta:2017hfi,Pratt:2018ebf,Pratt:2019fbj,An:2020vri,Du:2020bxp,Pratt:2021xvg,Savchuk:2023yeh}.

The importance of baryon conservation effects for controlling the non-critical baseline for proton number cumulants is now well-established~\cite{Pratt:2020ekp,Braun-Munzinger:2020jbk,Vovchenko:2021kxx}.
Canonical treatment of global baryon conservation provides the minimum baseline.
At the same time, even stronger, \emph{local} effects of exact charge conservation~\cite{Sakaida:2014pya,Vovchenko:2019kes,Braun-Munzinger:2023gsd} are likely to be relevant in heavy-ion collisions due to the presence of causally disconnected regions of the fireball~\cite{Castorina:2013mba}.
Controlling these effects is crucial for extracting physics relevant to the QCD phase structure from experimental measurements of fluctuations.
Currently, a simplified model of a truncated fireball in a conservation~(correlation) volume $V_c$~\cite{Vovchenko:2019kes} is being used to constrain the local conservation of QCD charges at LHC conditions through measurements of various fluctuations and correlations of hadron numbers~\cite{ALICE:2022xiu,ALICE:2022xpf,ALICE:2024rnr}.

In Refs.~\cite{Vovchenko:2020tsr,Vovchenko:2020gne,Vovchenko:2021yen} a subensemble acceptance method~(SAM) was developed to correct cumulants measured inside a spatial subvolume for global charge conservation effects.
Here, this method is generalized to compute the $n$-point spatial density correlator to the desired order.
The resulting explicit form of the balancing contributions in equilibrium allows one to introduce the effect of local charge conservation, which can be constrained with experimental measurements.
As a first application, it is demonstrated how the available measurements of net proton variance constrain the rapidity range of local baryon conservation at the LHC.

\paragraph*{\bf Definitions.}

The primary quantity of interest here is the $n$-point spatial density correlator $\mathcal{C}_n$, defined as
\eq{
\mathcal{C}_n({\bf r}_1, \ldots, {\bf r}_n) \equiv \left\langle\prod_{i=1}^n \delta \rho ({\bf r}_i) \right\rangle_c, \qquad n \geq 2,
}
and $\mathcal{C}_1(\rr_1) = \rho(\rr_1)$.
Here $\delta \rho(\rr_i) = \rho(\rr_i) - \mean{\rho(\rr_i)}$ is the density fluctuation at coordinate $\rr_i$, and the notation $\mean{\ldots}_c$ indicates cumulants, which differ from central moments starting from 4th order.
Up to order four, the correlators read
\eq{
\mathcal{C}_2(\rr_1,\rr_2) = \mean{\delta \rho_1 \delta \rho_2 }, \quad \mathcal{C}_3(\rr_1,\rr_2,\rr_3) = \mean{\delta \rho_1 \delta \rho_2 \delta \rho_3}, 
}
and~\cite{Pratt:2019fbj}
\eq{
\mathcal{C}_4(\rr_1,\rr_2,\rr_3,\rr_4) = \mean{\delta \rho_1 \delta \rho_2 \delta \rho_3 \delta \rho_4} 
- \mean{\delta \rho_1 \delta \rho_2}\mean{\delta \rho_3\delta \rho_4} \nonumber \\
- \mean{\delta \rho_1 \delta \rho_3}\mean{\delta \rho_2\delta \rho_4}
- \mean{\delta \rho_1 \delta \rho_4}\mean{\delta \rho_2\delta \rho_3}.
}
Here $\delta \rho_i \equiv \delta \rho(\rr_i)$.
Integrating $\mathcal{C}_n$ over any subvolume $V_s$ therefore yields the value of the $n$th-order cumulant of the charge distribution inside $V_s$, i.e.
\eq{
\kappa_n[B_{V_s}] = \int_{\rr_1 \in V_s} d \rr_1 \ldots \int_{\rr_n \in V_s} d \rr_n \, \mathcal{C}_n(\{\rr_i\})~.
}
Integration over distinct volumes provides joint cumulants and quantifies correlations among the charge numbers in different spatial domains.

\paragraph*{\bf 2-point correlator.}

The 2-point correlator can be written as a sum of local and balancing parts,
\eq{\label{eq:C2gen}
\mathcal{C}_2(\rr_1,\rr_2) = \chi_2 \delta(\rr_1 - \rr_2) + \mathcal{C}_2'(\rr_1,\rr_2).
}
The local part corresponds to the susceptibility $\chi_2$.
The balancing part satisfies $\int \mathcal{C}_2'(\rr_1, \rr_2) d \rr_{1,2} = -\chi_2$ since the charge variance $\kappa_2[B]$ vanishes in the full volume.
The correlation from global charge conservation is expected to be long range~\cite{Bzdak:2017ltv}, entailing that $\mathcal{C}_2'(\rr_1,\rr_2)$ is uniform.
One can thus empirically infer $\mathcal{C}_2'(\rr_1,\rr_2) = -\chi_2/V$ and
\eq{\label{eq:C2empir}
\mathcal{C}_2(\rr_1,\rr_2) = \chi_2 \delta(\rr_1 - \rr_2) - \frac{\chi_2}{V}.
}

\begin{figure}[t]
  \centering
  \includegraphics[width=.49\textwidth]{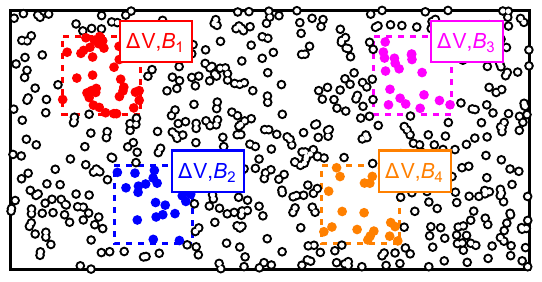}
  \caption{
  Schematic illustration of the partition of the system volume $V$ into distinct subvolumes of size $\Delta V$.
  The points illustrate the distribution of the conserved charge $B$ in a single microstate.
  }
  \label{fig:method}
\end{figure}

To obtain a rigorous derivation, extendable to high-order correlators, one can utilize the structure of the partition function in the thermodynamic limit.
First, split the total volume into $n$ distinct subvolumes of size $\Delta V$ plus the remainder volume equal to $V - n \Delta V$.
This is schematically illustrated in~Fig.~\ref{fig:method}, where different colored boxes show the subvolumes and the charge carriers inside them.
The subvolumes can exchange the conserved charge with the remainder volume, hence the dashed lines at their boundaries.
The subvolumes are assumed to be macroscopically large such as to capture the physics of correlation length, $\Delta V \gg \xi^3$.
Denoting the baryon numbers in each subvolume by $\{B_1, \ldots, B_{n}\}$, their joint probability is then proportional to the product of the canonical partition functions $Z(V,B)$~\footnote{The temperature variable $T$ has been dropped for brevity.} of the subvolumes~\cite{landau2013statistical}:
\eq{\label{eq:PBi}
P(\{B_i\}) & \propto \left[ \prod_{j=1}^n Z(\Delta V, B_j) \right] Z(V - n\Delta V, B - \sum_{j=1}^n B_j) \nonumber \\
& \propto  \left[ \prod_{j=1}^n e^{-\Delta V 
 f(\rho_j) } \right] e^{-(V-n\Delta V) f(\rho_{n+1})}
}
Here $f(\rho)$ is the free energy density, $\rho_j = B_j / \Delta V$ for $j = 1,\ldots,n$ and $\rho_{n+1} = (B - \sum_{j=1}^n B_j) / (V - n \Delta V)$.
The joint probability~\eqref{eq:PBi} can be used to calculate the joint cumulants $\mean{\delta B_1^{k_1} \ldots \delta B_n^{k_n}}_c$ and express these in terms of the susceptibilities $\chi_n$ that are encoded in the free energy density $f(\rho)$.
The algorithm has been presented in Refs.~\cite{Vovchenko:2020tsr,Vovchenko:2020gne} where a partition into two subvolumes was performed and the maximum term method applicable in the thermodynamic limit was used.
The generalization to $n+1$ subvolumes is straightforward and
the technical details are available in the \supmat.

For the second-order cumulants, one obtains
~[Eq.~\eqref{eq:k2app} in \supmat]
\eq{
\label{eq:k2}
\mean{\delta B_i \delta B_j} = \cc{2} \Delta V \delta_{ij} - \cc{2} \frac{(\Delta V)^2}{V}, \quad i,j = 1,\ldots,n.
}
The first term corresponds to coinciding 
subvolumes and can be identified as the local term proportional to $\delta(\rr_1 - \rr_2)$ in Eq.~\eqref{eq:C2gen}.
The second term corresponds to two distinct subvolumes and thus the non-local second term in Eq.~\eqref{eq:C2gen}.
The 2-point correlator, therefore, is
\eq{\label{eq:C2}
\mathcal{C}_2(\rr_1,\rr_2) = \chi_2 \delta(\rr_1 - \rr_2) - \frac{\chi_2}{V},
}
in agreement with the empirical derivation~[Eq.~\eqref{eq:C2empir}].

\paragraph*{\bf 3-point correlator.}

The three-point correlator, $\mathcal{C}_3(\rr_1,\rr_2,\rr_3) = \mean{\delta \rho_1 \delta \rho_2 \delta \rho_3}$, has three kinds of terms: (i) when all three coordinates coincide, $\rr_1 = \rr_2 = \rr_3$, (ii) when two of the coordinates coincide, e.g. $\rr_1 = \rr_2 \neq \rr_3$, and (iii) when all three coordinates are distinct $\rr_1 \neq \rr_2 \neq \rr_3$.
Thus, in addition to local and global terms, there are terms with a mixture of local correlations and balancing due to charge conservation.
The third-order joint cumulants read
~[Eq.~\eqref{eq:k3app} in \supmat]
\eq{\label{eq:rho3}
\mean{\delta B_i \delta B_j \delta B_k} & = \delta_{ijk} \chi_3 \Delta V - (\delta_{ij} + \delta_{ik} + \delta_{kj}) \chi_3 (\Delta V)^2 \nonumber \\
&  \quad + 2 \chi_3 (\Delta V)^3, \qquad i,j,k = 1,\ldots,n.
}
These are the terms contributing to the three-point correlator, $\mathcal{C}_3$.
In particular, the middle term in Eq.~\eqref{eq:rho3} corresponds to the cases where two of the three subvolumes coincide.
The correlator $C_3(\rr_1,\rr_2,\rr_3)$ has three such terms: (i) $\rr_1 = \rr_2 \neq \rr_3$, (ii) $\rr_1 = \rr_3 \neq \rr_2$, (iii) $\rr_2 = \rr_3 \neq \rr_1$.
Therefore,
\eq{\label{eq:C3}
\mathcal{C}_3(\rr_1,\rr_2,\rr_3) = \cc{3} \delta_{1,2,3} - \frac{\cc{3}}{V}[\delta_{1,2} + \delta_{1,3} + \delta_{2,3} ] + 2 \frac{\cc{3}}{V^2}~.
}
Here $\delta_{1,\ldots,n} = \prod_{i=2}^n \delta(\rr_1 - \rr_i)$.

\paragraph*{\bf 4-point correlator.}
\begin{widetext}
Following the iterative procedure to compute 4th-order joint cumulants one obtains~[Eq.~\eqref{eq:k4app} in \supmat]
\eq{\label{eq:C4}
\mathcal{C}_4(\rr_1,\rr_2,\rr_3,\rr_4) & = \cc{4} \delta_{1,2,3,4} - \frac{\cc{4}}{V}[\delta_{1,2,3} + \delta_{1,2,4} + \delta_{1,3,4} + \delta_{2,3,4} ] - \frac{(\cc{3})^2}{\cc{2} V} [\delta_{1,2} \delta_{3,4} + \delta_{1,3} \delta_{2,4} + \delta_{1,4} \delta_{2,3}] \nonumber \\
& \quad + \frac{1}{V^2} \left[\cc{4} + \frac{(\cc{3})^2}{\cc{2}} \right] \left[\delta_{1,2} + \delta_{1,3} + \delta_{1,4} + \delta_{2,3} + \delta_{2,4} + \delta_{3,4}\right] - \frac{3}{V^3} \left[\cc{4} + \frac{(\cc{3})^2}{\cc{2}} \right]~.
}
\end{widetext}

\paragraph{\bf Local conservation.}
Having the explicit expressions for the balancing terms entering the $n$-point density correlators allows one to introduce the local charge conservation effect by modulating the corresponding balancing contribution.
Here the considerations will be restricted to the 2-point correlator, which reads
\eq{\label{eq:C2local}
\mathcal{C}_2(\rr_1, \rr_2) = \chi_2 \left[
\delta(\rr_1 - \rr_2) - \frac{\varkappa(\rr_1,\rr_2)}{V}
\right].
}
Here $\varkappa(\rr_1,\rr_2)$ is the function that implements local charge conservation.
Ideally, the function $\varkappa(\rr_1,\rr_2)$ should be computed by solving a formidable task of propagating density fluctuations on a hydrodynamic background.
One would generally expect  $\varkappa(\rr_1,\rr_2)$ to be a localized function of the difference $|\rr_1 - \rr_2|$, such as a Gaussian.
It should be symmetric under exchange of its arguments, $\varkappa(\rr_1,\rr_2) = \varkappa(\rr_2,\rr_1)$, and satisfy the sum rules, $\int d {\rr_1} \varkappa(\rr_1,\rr_2) = \int d {\rr_2} \varkappa(\rr_1,\rr_2) = V$.
The case $\varkappa(\rr_1,\rr_2) = 1$ corresponds to the equilibrium limit of global conservation.
The grand-canonical limit is obtained by setting $\varkappa(\rr_1,\rr_2) = 0$.

\paragraph{\bf Net baryon number fluctuations at LHC.}

As a first application of the method, fluctuations of the net baryon number in Pb-Pb collisions at $\sNN = 5.02$ GeV are considered here.
Boost invariance is a good approximation of the midrapidity region where experimental measurements are performed at this energy. Hence, one can apply the developed formalism for uniform systems most straightforwardly to heavy-ion collisions at the LHC.
Only the longitudinal (spatial) rapidity coordinate $\eta$ is used, $\rr = \eta$, and the longitudinal boost invariance with a cut-off, $\eta \in [-\eta_{\rm max}, \eta_{\rm max}]$, is imposed.
The value of $\eta_{\rm max}$ is fixed such that the total fireball volume $V_{\rm tot} = 2 \eta_{\rm max} dV/dy$ is in line with the measurements of rapidity density of charged multiplicity~\cite{ALICE:2016fbt}. 
The latter imply a Gaussian distribution of the volume with a width $\sigma_V = 4.12 \pm 0.10$, giving $V_{\rm tot} = dV/dy \int dy \, e^{-\frac{y^2}{2\sigma_V^2}} = dV/dy \, \sqrt{2\pi} \sigma_V \simeq 10.2 \, dV/dy$, corresponding to $\eta_{\rm max} = 5.1$.

Local conservation is taken to be the only source of correlations, thus, the (grand-canonical) susceptibilities follow the Skellam distribution, $\chi_2 \propto d \mean{B + \bar{B}}/d \eta$.
The local baryon conservation in rapidity is modeled by a Gaussian,
\eq{\label{eq:LocalGauss}
\varkappa (\eta_1,\eta_2) = 2 \eta_{\rm max} \frac{\exp \left[-\frac{(\eta_1-\eta_2)^2}{2 \sigma_{y}^2}\right]}{\sqrt{2\pi} \sigma_{y} {\rm erf } \left(\frac{\eta_{\rm max}}{\sqrt{2} \sigma_{y}}\right)}.
}
To preserve the symmetry $\varkappa(\eta_1,\eta_2) = \varkappa(\eta_2,\eta_1)$ and the sum rule $\int d \eta_{1,2} \varkappa(\eta_1,\eta_2) = 2 \eta_{\rm max}$, periodic boundary conditions in $\eta$ are implemented such that $(\eta_1 - \eta_2)$ in Eq.~\eqref{eq:LocalGauss} ranges from $-\eta_{\rm max}$ to $\eta_{\rm max}$~\footnote{Periodic boundary conditions is an artefact of assuming perfectly Gaussian rapidity correlation in a system with a finite longitudinal extent. This artefact is negligible for observables near midrapidity.}.
For $\sigma_y \to \infty$, one recovers the global baryon conservation limit.

\begin{figure*}[t]
  \centering
  \includegraphics[width=.48\textwidth]{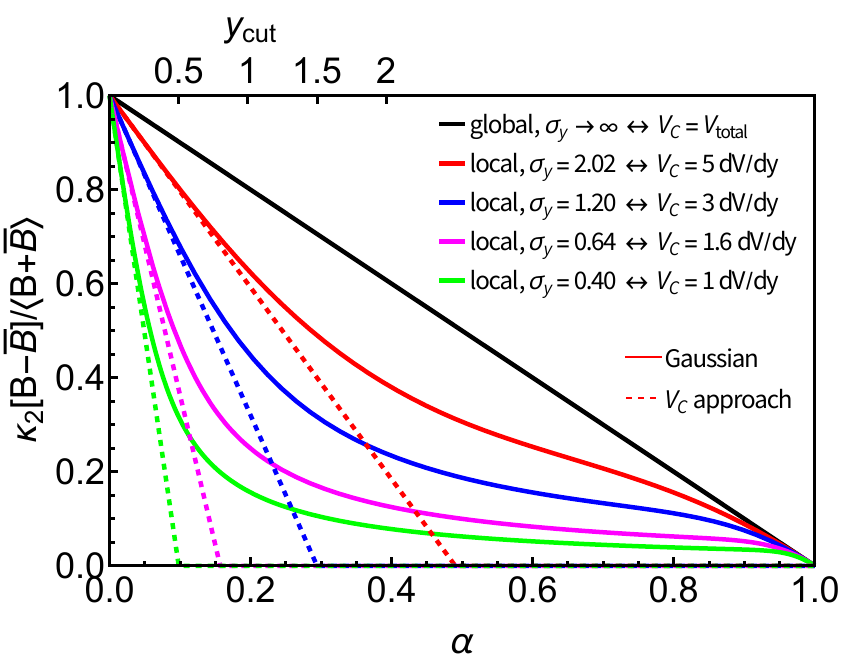}
  \includegraphics[width=.49\textwidth]{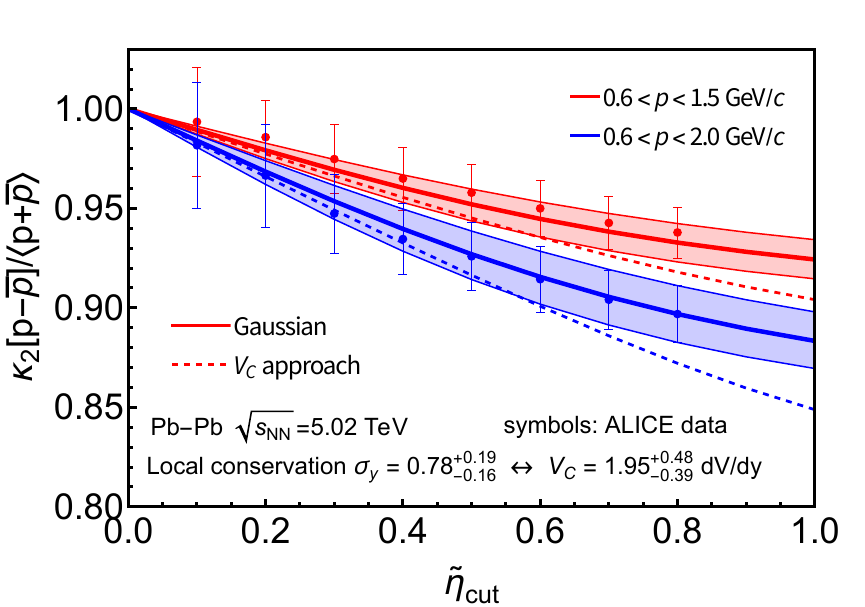}
  \caption{
  Left panel: The normalized variance of net-baryon distribution as a function of acceptance fraction $\alpha$ in spatial rapidity, $|\eta| < \eta_{\rm cut}$, computed for LHC conditions~($\sNN = 5.02$~TeV). The solid lines correspond to calculations with different values of Gaussian width $\sigma_y$ of local conservation in rapidity. 
  Right panel:
  Dependence of the normalized variance of net-proton distribution in 0-5\% central Pb-Pb collisions as a function of pseudorapidity cut. Black and red lines with a band depict the results in momentum acceptances of $0.6 < p < 1.5$~GeV/$c$ and $0.6 < p < 2.0$~GeV/$c$, respectively.
  The symbols show the experimental data of the ALICE Collaboration~\cite{ALICE:2022xpf}.
  The dashed lines in both panels correspond to the results from the equivalent $V_c$ approach.
  }
  \label{fig:ALICElocal}
\end{figure*}

The variance of net baryon number inside a (spatial) rapidity cut $|\eta| < \eta_{\rm cut}$ is obtained by integrating the 2-point density correlator, $\kappa_2[B-\bar{B}]|_{|\eta| < \eta_{\rm cut}} = \int_{-\eta_{\rm cut}}^{\eta_{\rm cut}} d \eta_1 \int_{-\eta_{\rm cut}}^{\eta_{\rm cut}} d \eta_2 \, \mathcal{C}_2(\eta_1,\eta_2)$.
One obtains
\begin{widetext}
\eq{\label{eq:k2Gauss}
\kappa_2[B-\bar{B}]|_{|\eta| < \eta_{\rm cut}}
& = \int_{-\eta_{\rm cut}}^{\eta_{\rm cut}} d \eta_1 \int_{-\eta_{\rm cut}}^{\eta_{\rm cut}} d \eta_2 \frac{d \mean{B+\bar{B}}}{d \eta} \left[ \delta(\eta_1 - \eta_2) - \frac{\exp \left[-\frac{(\eta_1-\eta_2)^2}{2 \sigma_{y}^2}\right]}{\sqrt{2\pi} \sigma_{y} \rm erf \left(\frac{\eta_{\rm max}}{\sqrt{2} \sigma_{y}}\right)} \right] \nonumber \\
& = \mean{B + \bar{B}} \left[ 1 - \frac{\left( 1 - e^{-\frac{2 \eta_{\rm cut}^2}{\sigma_{y}^2}}\right) \sigma_{y} + \eta_{\rm cut} \sqrt{2\pi} \rm erf \left( \frac{\sqrt{2} \eta_{\rm cut}}{\sigma_{y}} \right)}{\eta_{\rm cut} \sqrt{2\pi} \, {\rm erf} \left(\frac{\eta_{\rm max}}{\sqrt{2} \sigma_{y}}\right)} \right], \qquad  \eta_{\rm cut} < \eta_{\rm max} / 2 \nonumber \\
& = \mean{B + \bar{B}} \left[1 - \frac{\sqrt{2}\eta_{\rm max}}{\sqrt{\pi}\sigma_{y} {\rm erf} \left( \frac{\eta_{\rm max}}{\sqrt{2} \sigma_{y}} \right)} \alpha + O(\alpha^3) \right].
}   
\end{widetext}
Here $\alpha = \eta_{\rm cut} / \eta_{\rm max}$ is the fraction of the system inside the rapidity cut, and the expressions are obtained for $\eta_{\rm cut} < \eta_{\rm max} / 2$ where periodic boundary conditions are irrelevant.
The numerical calculations below are done for the full range, including $\eta_{\rm cut} > \eta_{\rm max} / 2$ where periodic boundary conditions are faithfully implemented.

Figure~\ref{fig:ALICElocal} shows the dependence of $\kappa_2[B-\bar{B}]/\mean{B+\bar{B}}$ on the acceptance fraction $\alpha$ for different values of $\sigma_y$.
In the limit $\sigma_{y} \to \infty$~(global baryon conservation) one obtains the well-known result, $\kappa_2[B-\bar{B}]/\mean{B+\bar{B}} = 1 - \alpha$~\cite{Bzdak:2012an,Pruneau:2019baa,Braun-Munzinger:2020jbk}.
With a decreasing value of $\sigma_y$, the drop with $\alpha$ is steeper.
At small $\alpha$, this is evident from Eq.~\eqref{eq:k2Gauss} by a steeper negative slope.
These results are qualitatively consistent with prior estimates of local baryon conservation~\cite{Sakaida:2014pya,Braun-Munzinger:2023gsd}.

\paragraph{\bf Connection to the $V_c$ approach.}
Analysis of the small $\alpha$ behavior allows one to establish the connection between Gaussian local conservation and the $V_c$ approach of Refs.~\cite{Vovchenko:2019kes,ALICE:2022xiu,ALICE:2022xpf,ALICE:2024rnr}.
In the $V_c$ approach, the fireball is truncated to $k$ units around midrapidity, $|\eta| < k/2$, which is treated in the canonical ensemble.
The $V_c$ approach is thus equivalent to a global conservation model with a reduced rapidity cut-off, $\eta_{\rm cut} \to k/2$, and thus entails a linear dependence on $\alpha$ with a modified slope
\eq{
\left(\frac{\kappa_2[B - \bar{B}]}{\mean{B + \bar{B}}}\right)_{V_c = k dV/d\eta} = 1 - \tilde \alpha = 1 - \frac{2 \eta_{\rm max}}{k} \alpha~.
}
Comparing it with Eq.~\eqref{eq:k2Gauss} one obtains the connection between $k$ and $\sigma_y$:
\eq{\label{eq:kVsVc}
k(\sigma_{y}) = \sqrt{2\pi} \sigma_{y} {\rm erf} \left( \frac{\eta_{\rm max}}{\sqrt{2} \sigma_y} \right) \approx \sqrt{2\pi} \sigma_y,
}
where the latter approximation applies for $\sigma_{y} \ll \eta_{\rm max}$.
The dashed lines in Fig.~\ref{fig:ALICElocal} depict the $V_c$ approach results using $k$ from Eq.~\eqref{eq:kVsVc}.
The comparison with the full lines allows one to establish the validity range of the $V_c$ approach.
The $V_c$ approach is accurate for $|y_{\rm cut}| \lesssim k/4$ and breaks down at higher rapidities.
For measurements within one unit at midrapidity one thus has to restrict $k \gtrsim 2$ when applying the $V_c$ approach.

\paragraph{\bf Proton fluctuations and kinematical cuts.}

Momentum cuts replace the coordinate cuts in experimental measurements.
In particular, net proton cumulants have been measured by the ALICE Collaboration in Pb-Pb collisions at the LHC for utilizing cuts in 3-momentum, $0.6 < p < 1.5(2.0)$~GeV/$c$, and kinematical pseudorapidity, $|\tilde \eta| < \tilde \eta_{\rm cut}$~\cite{ALICE:2019nbs,ALICE:2022xpf}.
The present formalism can be used to incorporate such kinematical cuts if the acceptance probabilities for all (anti)baryons are independent.
Namely, one obtains the net proton variance by integrating $\mathcal{C}_2(\eta_1,\eta_2)$ over all spatial rapidities and applying a Bernoulli filter for each $(\eta_1,\eta_2)$ pair to account for momentum cuts.

Let us denote by $p(\eta)$ the probability that an (anti)baryon emitted from spatial rapidity $\eta$ ends up as an (anti)proton inside the acceptance, taken to be identical for protons and antiprotons.
The first term in $\mathcal{C}_2$~\eqref{eq:C2local} is the self-correlation term~\footnote{This is the case in the ideal gas limit. In case of an interacting system the first term in Eq.~\eqref{eq:C2local} contains both the self-correlation and local two-baryon correlations.}, thus the binomial acceptance dilutes it by a single power of $p(\eta)$. 
The second term corresponds to the balancing contribution and thus the the two-baryon correlations. It is multiplied by $p(\eta_1) p(\eta_2)$~\cite{Kitazawa:2012at,Bzdak:2012ab}.
The normalized variance of the net proton number inside momentum acceptance thus reads
\eq{\label{eq:k2p}
\frac{\kappa_2[\rm p-\bar{\rm p}]_{\rm acc}}{\mean{\rm p+\bar{\rm p}}} = 1 - \frac{\mean{p(\eta_1) p(\eta_2)}}{\mean{p(\eta)}},
}
with $\mean{p(\eta)} = \frac{1}{2 \eta_{\rm max}} \int d \eta p(\eta)$ and $\mean{p(\eta_1) p(\eta_2)} = \frac{1}{(2 \eta_{\rm max})^2} \int d \eta_1 d \eta_2 p(\eta_1) p(\eta_2) \varkappa(\eta_1,\eta_2)$.

$p(\eta)$ is evaluated using the blast-wave model~\cite{Schnedermann:1993ws}, which reasonably describes proton transverse momentum spectra~\cite{ALICE:2019hno}.
The details are available in \supmat.
In addition to the kinematic cuts, $p(\eta)$ also contains efficiency factor $q \simeq 0.33$ to distinguish protons from all baryons.
The value of $q$ is based on the HRG model estimation at freeze-out~\cite{Vovchenko:2019pjl}.

The results are shown in the right panel of Fig.~\ref{fig:ALICElocal} and they exhibit sensitivity to the range $\sigma_y$ of local baryon conservation.
By fitting the largest acceptance point, $0.6 < p < 2.0$~GeV/$c$ and $|\tilde \eta| < 0.8$, one obtains $\sigma_{y} = 0.78^{+0.19}_{-0.16}$. This corresponds to the conservation volume $V_c = (1.95^{+0.48}_{-0.39}) dV/dy$.
With the same parameters, one obtains a reasonable description of the data for smaller pseudorapidity cuts as well as for the momentum cut $0.6 < p < 1.5$~GeV/$c$.
Note that this estimate neglects possible baryon interactions~\cite{Vovchenko:2020kwg} and baryon annihilation~\cite{Savchuk:2021aog}.
Results of the equivalent $V_c$ approach are also shown in Fig.~\ref{fig:ALICElocal}, by the dashed lines.
For the extracted values of $\sigma_y$ one observes that the $V_c$ approach overpredicts the suppression due to local baryon conservation.
In particular, the same data analyzed in the $V_c$ approach suggests $V_c \simeq 3 dV/dy$~\cite{ALICE:2024rnr}, which is considerably larger than given by the Gaussian charge correlation.
Therefore, estimates of the correlation volume based on $V_c$ should be regarded as an upper bound on the value of $k$.

The conservation volume $V_c = (1.95^{+0.48}_{-0.39}) dV/dy$ is about five times smaller than the total fireball volume $V_{\rm tot} \simeq 10.2 dV/dy$.
Thus, experimental measurements of the net proton variance at $\sNN = 5.02$~TeV Pb-Pb collisions provide significant evidence for local baryon conservation, as for global baryon conservation only one would expect $V_c = V_{\rm tot}$.
This observation calls for a more detailed analysis of local conservation effects in other systems, such as Au-Au collisions at RHIC-BES, where so far only global baryon conservation has been considered.
The analysis of RHIC-BES-I data in~\cite{Vovchenko:2021kxx} indicates the additional suppression of proton number variance relative to global baryon conservation only at $\sNN \gtrsim 20$~GeV, which was modeled by the excluded volume effect in~\cite{Vovchenko:2021kxx}.
This suppression may also be indicative of local baryon conservation effects at RHIC, similar to what is obtained for LHC in the present work.
These questions will be addressed in a forthcoming study utilizing the extension of the formalism to non-boost-invariant systems.

\paragraph{\bf Summary and outlook.}

This work presented a novel formalism to compute the $n$-point correlator of a conserved charge density in the canonical ensemble.
The main results are given by Eqs.~\eqref{eq:C2},~\eqref{eq:C3}, and~\eqref{eq:C4} expressing 2-,3-, and 4-point density correlations in terms of grand-canonical susceptibilities.
The results are universal and applicable for any system in equilibrium in the thermodynamic limit, namely when $V \gg \xi^3$ holds.
By modulating the balancing contribution term to the 2-point baryon density correlator, a Gaussian local conservation of baryon charge was introduced.
The analysis of experimental data on net proton fluctuations from 5.02~TeV Pb-Pb collisions indicates local baryon conservation with a Gaussian width $\sigma_{y} = 0.78^{+0.19}_{-0.16}$, corresponding to the conservation volume $V_c$ spanning $k \simeq \sqrt{2\pi} \sigma_y = (1.95^{+0.48}_{-0.39})$ units of rapidity.

The formalism leaves plenty of room for extensions and applications.
One possibility is to utilize the input for the susceptibilities from lattice QCD or effective theories, or incorporate local correlations among baryons such as the repulsive excluded volume effect, baryon clustering, and baryon annihilation.
One can also extend it to multiple conserved charges, as well as non-conserved quantities correlated to conserved charges, such as the individual hadron numbers and balance functions, as well as factorial cumulants~\cite{Barej:2022jij}.
This will allow one to apply the formalism at lower energies, such as RHIC, SPS, HADES, and CBM, in particular by utilizing particlization hypersurfaces from hydrodynamic simulations.
It is also of interest to implement the correlations induced by (local) charge conservation into a Monte Carlo sampler of the hadron resonance gas at particlization.


\begin{acknowledgments}

\emph{Acknowledgments.} 
The author acknowledges fruitful discussions with Volker Koch and Scott Pratt.

\end{acknowledgments}

\bibliography{bibliography}

\begin{appendix}
\widetext 
\setcounter{equation}{0}
\setcounter{figure}{0}
\renewcommand{\theequation}{A.\arabic{equation}}
\makeatletter
\section{\large \bf Supplemental material}
\section{Derivation of joint cumulants inside different subvolumes}
\label{app:Deriv}

The total volume $V$ of the thermal system in the canonical ensemble is partitioned into $n+1$ subvolumes: $n$ subvolumes of size $\Delta V$ and the remaining subvolume of size $V - n \Delta V$.
The total charge $B = \sum_{i=1}^{n+1} B_i$ does not fluctuate but the charges $\bvar B = (B_1,\ldots,B_n)$ inside the various subvolumes do.
Thermodynamic limit is assumed, $V \to \infty$, and $\Delta V / V = \rm const$.
Under these conditions, one can write the joint probability for the $\bvar B$ distribution as a product of the canonical partition functions of the subvolumes:
\eq{\label{eq:PBis}
P(\bvar B) & \propto \left[ \prod_{j=1}^n Z(\Delta V, B_j) \right] Z(V - n\Delta V, B - \sum_{j=1}^n B_j) 
\propto  \left[ \prod_{j=1}^n e^{-\Delta V 
 f(\rho_j) } \right] e^{-(V-n\Delta V) f(\rho_{n+1})}.
}
Here the partition functions are expressed in terms of the free energy density: $Z(V,B) \stackrel{V \to \infty}{\sim} e^{-V f(\rho)}$.
Here $\rho_j = B_j / \Delta V$ for $j = 1,\ldots,n$ and $\rho_{n+1} = (B - \sum_{j=1}^n B_j) / (V - n \Delta V)$.

The joint cumulant generating function for the $\bvar B$ distribution is defined as
\eq{
G_{\bvar B}(\bvar t) = \ln \mean{e^{\sum_{i=1}^n t_i B_i}} = \ln \left[ \sum_{\bvar B} \exp\left(\sum_{i=1}^n t_i B_i\right) P(\bvar B) \right]~,
}
such that the derivatives of $G_{\bvar B}(\bvar t)$ define the joint cumulants
\eq{\label{eq:jointB}
\mean{\delta B_1^{k_1} \ldots \delta B_n^{k_n}}_c = \left. \frac{\partial^{k_1 + \ldots +k_n} G_{\bvar B}(\bvar t)}{\partial t_1^{k_1} \ldots \partial t_n^{k_n}} \right|_{t_1 = \ldots = t_n = 0} .
}

The first-order cumulants correspond to first derivatives of $G_{\bvar B}(\bvar t)$.
One can define the first-order cumulant at finite values of $\bvar t$, namely
\eq{\label{eq:Bmeant}
\mean{B_i}(\bvar t) & = \left(\frac{\partial G_{\bvar B}(\bvar B)}{\partial t_i} \right)_{t_{j\neq i}}  = \frac{\displaystyle \sum_{\bvar B} B_i \tilde P (\bvar B; \bvar t)}{\displaystyle \sum_{\bvar B} \tilde P (\bvar B; \bvar t)},
}
with the (un-normalized) $\bvar t$-dependent  probability
\eq{
\tilde P (\bvar B; \bvar t) = \exp\left(\sum_{i=1}^n t_i B_i\right) P(\bvar B),
}

In the thermodynamic limit, $V \to \infty$,  $\tilde P$ has a sharp maximum at
the mean value of $\bvar B$, $\mean{\bvar B(\bvar t)}$. 
The location of this maximum is given by the equation $\partial \tilde P(\bvar B; \bvar t) / \partial \bvar B = 0$, resulting in a system of implicit equations determining  $\mean{\bvar B(\bvar t)}$:
\eq{\label{eq:ts}
t_i = \mu[\rho_i(\bvar t)] - \mu[\rho_{n+1}(\bvar t)], \quad i = 1,\ldots, n.
}
Here $\mu$ is the chemical potential~(defined as $\mu = \partial f(\rho) / \partial \rho$), $\rho_i(\bvar t) = \mean{B_i}(\bvar t) / \Delta V$ for $j = 1,\ldots,n$ and $\rho_{n+1}(\bvar t) = (B - \sum_{j=1}^n \mean{B_j(\bvar t)}) / (V - n \Delta V)$.

\vskip2pt

\paragraph{\bf Second order cumulants.}

Given the $\bvar t$-dependent first-order cumulants $\mean{B_i}(\bvar t)$, the second order-cumulants are 
\eq{\label{eq:k2app}
\mean{\delta B_i \delta B_j} (\bvar t) = \frac{\partial \mean{B_i} (\bvar t)}{\partial t_j}.
}
To evaluate $\mean{\delta B_i \delta B_j} (\bvar t)$ one differentiates Eq.~\eqref{eq:ts} with respect to $t_j$:
\eq{\label{eq:delt1}
\delta_{ij} = \frac{\partial \mu_i}{\partial \rho_i} \,  \frac{\partial \rho_i}{\partial \mean{B_i}} \,  \frac{\partial \mean{B_i}}{\partial t_j} - \frac{\partial \mu_{n+1}}{\partial \rho_{n+1}} \,  \sum_{k=1}^n \frac{\partial \rho_{n+1}}{\partial \mean{B_k}} \,  \frac{\partial \mean{B_k}}{\partial t_j}.
}
Here $\mu_i = \mu[\rho_i(\bvar t)]$.
One observes that
\eq{
\frac{\partial \rho_i}{\partial \mean{B_i}} = \frac{1}{\Delta V}, \qquad \frac{\partial \rho_{n+1}}{\partial \mean{B_k}} = -\frac{1}{V - n \Delta V},
}
and
\eq{
\frac{\partial \mu_i}{\partial \rho_i} = \frac{1}{\chi_{2,i}}, \qquad \frac{\partial \mean{B_k}}{\partial t_j} = \mean{\delta B_k \delta B_j},
}
with $\chi_{2,i}(\bvar t) = \chi_2[\rho_i(\bvar t)]$.
The second-order cumulants, therefore, satisfy the following system of linear equations
\eq{
\left[ \frac{\delta_{ik}}{\Delta V \chi_{2,i}} + \frac{1}{(V - n \Delta V) \chi_{2,n+1}} \right] \mean{\delta B_k \delta B_j} = \delta_{ij}.
}
This system of equations can be solved explicitly to give
\eq{\label{eq:Bij}
& \mean{\delta B_i \delta B_j} (\bvar t) = \Delta V \chi_{2,i} \left[ \delta_{ij}  - \frac{\Delta V \chi_{2,j}}{\Delta V (\sum_{k=1}^n \chi_{2,k}) + (V - n \Delta V) \chi_{2,n+1}} \right].
}

For $\bvar t = 0$ one has $\chi_{2,i} = \chi_{2,n+1} = \chi_2$ and thus
\eq{
\mean{\delta B_i \delta B_j} = \Delta V \chi_{2} \left[ \delta_{ij} - \frac{\Delta V}{V} \right].
}

\vskip2pt
\paragraph{\bf High-order cumulants.}

Third-order cumulants can be obtained by differentiating Eq.~\eqref{eq:Bij} with respect to $t_k$.
By definition
\eq{
\mean{\delta B_i \delta B_j \delta B_k} (\bvar t) = \frac{\mean{\delta B_i \delta B_j} (\bvar t)}{\partial t_k}. 
}
The resulting expressions are tedious but can be straightforwardly implemented using symbolic calculus systems such as \texttt{Mathematica} used in the present work.
At $\bvar t = 0$ one obtains
\eq{\label{eq:k3app}
\mean{\delta B_i \delta B_j \delta B_k} & = \delta_{ijk} \chi_3 \Delta V - (\delta_{ij} + \delta_{ik} + \delta_{kj}) \chi_3 (\Delta V)^2  + 2 \chi_3 (\Delta V)^3.
}

The fourth (and higher) order cumulants are obtained iteratively by differentiating the previously obtained $\bvar t$-dependent lower-order cumulants with respect to the components of $\bvar t$.
Fourth order cumulants for $\bvar t = 0$ read
\eq{\label{eq:k4app}
\mean{\delta B_i \delta B_j \delta B_k \delta B_l}_c & = \Delta V \cc{4} \delta_{ijkl} - \cc{4} \frac{(\Delta V)^2}{V}[\delta_{ijk} + \delta_{ijl} + \delta_{ikl} + \delta_{jkl} ] - \frac{(\cc{3})^2}{\cc{2}} \frac{(\Delta V)^2}{ V} [\delta_{ij} \delta_{kl} + \delta_{ik} \delta_{jl} + \delta_{il} \delta_{jk}] \nonumber \\
& \quad + \frac{(\Delta V)^3}{V^2} \left[\cc{4} + \frac{(\cc{3})^2}{\cc{2}} \right] \left[\delta_{ij} + \delta_{ik} + \delta_{il} + \delta_{jk} + \delta_{jl} + \delta_{kl}\right] - \frac{3 (\Delta V)^4}{V^3} \left[\cc{4} + \frac{(\cc{3})^2}{\cc{2}} \right]~.
}

\section{Momentum cuts}

The formalism allows one to calculate density correlations and fluctuations in the coordinate space. However, in the experiment, coordinates are integrated over, and momentum cuts are incorporated instead.
In principle, one would need to define a density correlator that is differential in both the coordinates and momenta simultaneously.
In the particular case of a binomial acceptance in momentum, namely when the acceptance probabilities of the measured particles are independent, one can implement kinematic cuts in a simpler way.

Consider the fluctuations of net baryon number at LHC conditions measured as the difference between baryons $B$ and antibaryons $\bar{B}$.
In the ideal gas limit, the leading two baryon number susceptibilities read
\eq{
\cc{1}^B = \frac{d \mean{B}}{d \eta} - \frac{d \mean{\bar{B}}}{d \eta}, \qquad \cc{2}^B = \frac{d \mean{B}}{d \eta} + \frac{d \mean{\bar{B}}}{d \eta},
}
and the two-point baryon density correlator is
\eq{\label{eq:C2Bmomentum}
\mathcal{C}_2^B(\eta_1,\eta_2) = \frac{d \mean{B + \bar{B}}}{d \eta} \delta(\eta_1 - \eta_2) - \frac{d \mean{B + \bar{B}}}{d \eta} \frac{\varkappa(\eta_1,\eta_2)}{2 \eta_{\rm max}}.
}

The first term in Eq.~\eqref{eq:C2Bmomentum} corresponds to the self-correlation of (anti)baryons among themselves at $\eta = \eta_1 = \eta_2$. 
The second term is the balancing contribution due to (local) baryon conservation which corresponds to two-baryon correlations.

Momentum cuts and efficiency dilute the correlation from each $(\eta_1,\eta_2)$ pair.
If $p(\eta)$ is the probability that a randomly emitted (anti)baryon from spatial rapidity $\eta$ ends up as an (anti)proton inside the acceptance, then the self-correlation term is diluted by a factor $p(\eta_1)$ and the two-particle correlation term is diluted by factor $p(\eta_1) p(\eta_2)$.
The variance of net proton number inside momentum acceptance reads
\eq{
\kappa_2[\rm p - \bar{\rm p}]_{\rm acc} & = \int_{-\eta_{\rm max}}^{\eta_{\rm max}} d \eta_1 \int_{-\eta_{\rm max}}^{\eta_{\rm max}} d \eta_2 \left[ p(\eta_1) \frac{d \mean{B + \bar{B}}}{d \eta} \delta(\eta_1 - \eta_2) - p(\eta_1) p(\eta_2) \frac{d \mean{B + \bar{B}}}{d \eta} \frac{\varkappa(\eta_1,\eta_2)}{2 \eta_{\rm max}}
\right] \nonumber \\
& = \mean{B + \bar{B}}_{\rm tot} \mean{p(\eta)}  - \mean{B + \bar{B}}_{\rm tot} \mean{p(\eta_1) p(\eta_2)},
}
where $\mean{B+\bar{B}}_{\rm tot} = 2 \eta_{\rm max} \frac{d \mean{B + \bar{B}}}{d \eta}$ and
\eq{
\mean{p(\eta)} & = \frac{1}{2 \eta_{\rm max}} \int_{-\etam}^{\etam} d \eta \, p(\eta), \\
\mean{p(\eta_1)p(\eta_2)} & = \frac{1}{(2 \etam)^2} \int_{-\etam}^{\etam} d \eta_1 \int_{-\etam}^{\etam} d \eta_2 \, p(\eta_1) p(\eta_2) \varkappa(\eta_1,\eta_2).
}
Given that $\mean{\rm p+\bar{\rm p}}_{\rm acc} = \mean{p(\eta)} \mean{B+\bar{B}}_{\rm tot}$ one obtains
\eq{
\frac{\kappa_2[\rm p-\bar{\rm p}]_{\rm acc}}{\mean{\rm p+\bar{\rm p}}} = 1 - \frac{\mean{p(\eta_1) p(\eta_2)}}{\mean{p(\eta)}}.
}

\section{Acceptance probability from the blast-wave model}

The acceptance probability reads
\eq{
p(\eta) = q \, p_{\rm cut} (\eta).
}
Here $q$ is the efficiency factor, which is the probability that (anti)baryon is an (anti)proton in the final state.
$p_{\rm cut} (\eta)$ is the kinematical probability that a baryon emitted from spatial rapidity $\eta$ ends up in the given momentum acceptance.
At LHC conditions, $q \approx 0.33$ based on HRG model calculations at the freeze-out temperature $T = 155$~MeV.
Both $q$ and $p(\eta)$ are taken to be identical for baryons and antibaryons.

To calculate $p(\eta)$ one can utilize the single-particle momentum distribution.
The blast-wave model gives the following momentum distribution of protons emitted from spatial rapidity $\eta$
\eq{\label{eq:BW1}
\frac{d^3N}{p_T dp_T dy} (\eta) & \propto m_T \cosh (y - \eta) \, \int_0^1 \zeta \, d \zeta  e^{-\frac{m_T \, \cosh \rho \, \cosh (y - \eta)}{T}} I_0 \left(\frac{p_T \sinh \rho}{T}\right),
}
where $\rho = \tanh^{-1} ( \beta_s \zeta^{n_{\rm BW}})$ and $m_T = \sqrt{p_T^2 + m_N^2}$.

The blast-wave parameters are $T = 90$~MeV, $\beta_s = 0.906$, and $n_{\rm BW} = 0.735$ are  for 0-5\% Pb-Pb collisions at $\sNN = 5.02$~TeV, taken from Ref.~\cite{ALICE:2019hno}.
Net proton number fluctuations have been measured in acceptance with cuts in 3-momentum, $p_{\rm min} < p < p_{\rm max}$, and pseudorapidity, $|\tilde \eta| < \tilde \eta_{\rm cut}$.
One can transform~\eqref{eq:BW1} into the distribution in $(p,\tilde \eta)$ variables through the variable change
\eq{
y = {\rm arcsinh} \left[ \frac{p}{\sqrt{m_N^2 (\cosh \tilde \eta)^2 + p^2}} \sinh \tilde \eta \right], \qquad
p_T = \frac{p}{\cosh(\tilde \eta)},
\qquad
d p_T dy = \frac{p}{\sqrt{p^2+m_N^2} \cosh \tilde \eta} dp d \tilde \eta,
}
giving
\eq{
\frac{d^3N}{dp d \tilde \eta} (\eta) & \propto \frac{p^2 m_T \cosh (y - \eta)}{(\cosh \tilde \eta)^2 \sqrt{m_N^2 + p^2}}  \, \int_0^1 \zeta \, d \zeta  e^{-\frac{m_T \, \cosh \rho \, \cosh (y - \eta)}{T}} I_0 \left(\frac{p_T \sinh \rho}{T}\right),
}
and
\eq{
p_{\rm cut}(\eta)  = \frac{\int_{p_{\rm min}}^{p_{\rm max}} dp \int_{-\tilde \eta_{\rm cut}/2}^{\tilde \eta_{\rm cut}/2} d \tilde \eta \frac{d^3N}{dp d \tilde \eta} (\eta)}{\int_{0}^{\infty} dp \int_{-\infty}^{\infty} d \tilde \eta \frac{d^3N}{dp d \tilde \eta} (\eta)}~.
}

\end{appendix}


\end{document}